\documentclass[aps,prd,showpacs,twocolumn,floatfix,nofootinbib,
superscriptaddress]{revtex4-2}

\usepackage[dvipsnames,svgnames]{xcolor}

\usepackage{amsmath,amssymb}
\usepackage{hyperref}

\usepackage{slashed}
\usepackage{amsfonts}
\usepackage{graphicx}
\usepackage{amsmath}
\usepackage{lineno}
\usepackage{braket}
\usepackage[utf8]{inputenc}
\usepackage{babel}
\usepackage{nicefrac}
\usepackage{graphicx}
\usepackage[normalem]{ulem}
\usepackage{pstricks}
\usepackage{xcolor}
\usepackage{url}
\usepackage{empheq}  
\usepackage{hyperref}
\usepackage[normalem]{ulem}
\usepackage[nameinlink, capitalise, noabbrev]{cleveref}

\def\sss{\scriptscriptstyle}
\def\nn{\nonumber}
\def\CP{$C\!P$~}

\def\mev{\ensuremath{\mathrm{Me\kern -0.1em V}}}
\def\gev{\ensuremath{\mathrm{Ge\kern -0.1em V}}}
\newcommand{\modulus}[1]{\ensuremath{\left| #1 \right|}}

\allowdisplaybreaks

\begin{document}

\title{Implications of the evidence for direct  $\mathbf{CP}$ violation in 
$D\to \pi^+\pi^-$ decays}

\author{Rahul Sinha}%
\email{rsinha@hawaii.edu}%
\email{sinha@imsc.res.in}%
\affiliation{Department of Physics and Astronomy, University of
Hawaii, Honolulu, HI 96822, USA}%
\affiliation{The Institute of Mathematical Sciences, Taramani, Chennai
600113, India}%

\author{Thomas E. Browder}%
\email{teb@phys.hawaii.edu}%
\affiliation{Department of Physics and Astronomy, University of
Hawaii, Honolulu, HI 96822, USA}%

\author{N.~G.~Deshpande}%
\email{desh@uoregon.edu}%
\affiliation{Institute for Fundamental Science, University of Oregon,
Eugene, Oregon 94703, USA}%

\author{Dibyakrupa Sahoo}%
\email{Dibyakrupa.Sahoo@fuw.edu.pl}%
\affiliation{Faculty of Physics, Institute of Theoretical Physics,
University of Warsaw, ul. Pasteura 5, 02-093 Warsaw, Poland}%

\author{Nita Sinha}%
\email{nita@imsc.res.in}%
\affiliation{The Institute of Mathematical Sciences, Taramani, Chennai
600113, India}%

\date{\today}

\begin{abstract}
The observation of \CP violation in the difference of \CP asymmetries
between $D\to K^+K^-$ and $D\to \pi^+\pi^-$ has raised a debate
whether the observed asymmetries can be regarded as a signal of
physics beyond the standard model (SM). In this paper we obtain all
the topological amplitudes and isospin amplitudes directly from
measured observables for $D\to \pi\pi$. These results unambiguously
imply a very large penguin contribution, having a central value $4.74$
times the magnitude of the amplitude for $D^0\to\pi^+ \pi^-$. This
fitted central value differs from a reasonable SM estimate of $10\%$
with a significance greater than $3.3\sigma$.  In contrast to previous
studies, we present model-independent arguments based only on
unitarity of re-scattering amplitudes to show that large penguins
cannot arise from re-scattering alone and likely indicate physics
beyond the SM. In a model-independent approach we show how a very
small contribution from physics beyond the SM with a large weak phase
alleviates the problem.
\end{abstract}

\maketitle 

\section{Introduction}

The observation of \CP violation in $D$ meson decays has long been
regarded as a smoking gun signal for new physics (NP) or physics
beyond the SM. However, the observation of \CP violation in singly
Cabibbo suppressed $D$ meson decays~\cite{LHCb:2019hro,LHCb:2022lry}
has raised a debate whether the observed asymmetries can be regarded
as a signal of NP. The confusion stems from the difficultly  in
reliably estimating  long distance contributions and also convincingly
establishing that the penguin contribution is indeed too large to be
acceptable within the SM. Several studies~\cite{Bause:2022jes,
Buccella:2019kpn, Li:2019hho, Falk:2001hx, Grossman:2019xcj,
Bediaga:2022sxw, Pich:2023kim, Gavrilova:2023fzy, Franco:2012ck,
Brod:2012ud, Hiller:2012xm, Cheng:2012wr, ValeSilva:2024vmv} have
examined the issues indicating the likelihood that the observed
asymmetries may arise from NP. In this paper we estimate the size of
the penguin and tree contributions as well as isospin contributions
directly from experimental data. We find that the penguin
contributions are unexpectedly large, having a central value $4.74$
times the magnitude of the amplitude for $D^0\to\pi^+ \pi^-$.  This
fitted central value differs from a reasonable SM estimate of $10\%$
with a significance greater than $3.3\sigma$. We argue that the
evidence for \CP violation in $D\to \pi^+\pi^-$ is a signal of NP at
greater than $3.3\sigma$. We also examine final state interactions in
a completely model-independent way based only on unitarity in contrast
to previous studies. Our conclusion depends largely on the measured
value of the direct \CP asymmetry in $D\to\pi^+\pi^-$. In a
model-independent approach we show how a very small contribution from
NP with a large weak phase alleviates the problem. If this evidence is
indeed a signal of NP, large CP violation must show up in other modes
and experiments. Search for other signals of \CP violation in $D$
decays are essential to substantiate the evidence of NP.

\section{Analysis of \texorpdfstring{$\boldsymbol{D \to \pi \pi}$}{D --> pi pi} amplitudes}

The amplitudes for  $D\to \pi\pi$ decay modes can be written assuming
isospin as
\begin{align}
\label{eq:isospin_rels} A(D^0\to\pi^+\pi^-)&= \sqrt{2}
\Big(\!\lambda_d (T\!+\!E^d)\!+\!\lambda_i
(P^i\!+\!P\!E^i\!+\!P\!A^i)\!\Big), \nn\\%
A(D^0\to\pi^0\pi^0)&= \Big(\!\lambda_d (C\!-\!E^d)\!-\!\lambda_i
(P^i\!+\!P\!E^i\!+\!P\!A^i)\!\Big), \nn\\%
A(D^+\to\pi^+\pi^0)&= \lambda_d (T+C),
\end{align}
where $\lambda_i=V_{ci}^* V_{ui}$ and it is assumed that $i$ is summed
over $d$, $s$ and $b$ quarks. $T$, $C$, $E$, $P$, $P\!E$ and $P\!A$
are color-allowed external $W$-emission, color-suppressed internal
$W$-emission, $W$-exchange, penguin, penguin-exchange and
penguin-annihilation topological amplitudes, respectively. Unitarity
of the CKM matrix requires that $\lambda_d+\lambda_s+\lambda_b=0$,
which can be used to eliminate $\lambda_b$ in
Eq.~\eqref{eq:isospin_rels}. We choose this parametrization, resulting
in more accurate estimates. Within the SM, $\lambda_{d,s}$ can be
defined within the Wolfenstein parametrization up to ${\cal
O}(\lambda^6)$ as
\begin{align}
\lambda_d \!=& \Big(\!-\lambda+\tfrac{1}{2}A^2
\lambda^5\big(1-2(\rho-i\eta)\big)\Big) \!
\Big(1-\tfrac{1}{2}\lambda^2-\tfrac{1}{8}\lambda^4  \Big)\nn \\%
 =& (0.219113\pm 0.000625) e^{i(-0.000633\pm 0.000026)}, \\%
\lambda_s\! =&
\lambda\Big(\!1-\tfrac{1}{2}\lambda^2-\tfrac{1}{8}\lambda^4
\big(1+4A^2\big)\Big) \!%nn \\%
=\! 0.219045\pm 0.000624,\nn
\end{align}
$\lambda=0.22501\pm 0.00068$, $A=0.826^{+0.016}_{-0.015}$,
$(\bar{\rho},\bar{\eta})=(\rho,\eta)\times {\scriptsize
(1\!-\!\lambda^2/2\!+\!\cdots)}$, $\bar{\rho}\! =\! 0.1591\pm 0.0094$,
$\bar{\eta} \!=\!
0.3523^{+0.0073}_{-0.0071}$~\cite{ParticleDataGroup:2024cfk}.

The decay amplitudes in Eq.~\eqref{eq:isospin_rels} can be recast as,
\begin{align}
\label{eq:isospin_rels_simp} A(D^0\to\pi^+\pi^-)\equiv & A^{+-}=
\sqrt{2} ( t + e^{i\phi} p),  \nn\\%
A(D^0\to\pi^0\pi^0)\equiv & A^{00}=  c- e^{i\phi} p, \nn\\%
A(D^+\to\pi^+\pi^0)\equiv & A^{+0}=  (t+c),
\end{align}
where within the SM the weak phase
$\phi\equiv\arg(-{\lambda_s}/{\lambda_d})$ is $\phi=(0.000633\pm
0.000026)= (0.0363\pm 0.0015)^\circ$, and
\begin{align}
\label{eq:topol_amps} t = & \left|\lambda_d\right|
(T+E^d+P^{db}+P\!E^{db}+P\!A^{db}), \nn\\%
c = & \left|\lambda_d\right| (C-E^d-P^{db}-P\!E^{db}-P\!A^{db}),
\nn\\%
p = & \left|\lambda_s\right| (P^{sb}+P\!E^{sb}+P\!A^{sb}),
\end{align}
with $P^{ij}\equiv P^i-P^j$, $P\!E^{ij}\equiv P\!E^{i}-P\!E^{j}$ and
$P\!A^{ij}\equiv P\!A^{i}-P\!A^{j}$.  It may be noted that $t$, $c$
and $p$ are defined to include both the magnitude and any relevant
strong phase. Amplitudes for the conjugate modes
$A(\bar{D}^0\to\pi^+\pi^-)\equiv \bar{A}^{+-}$,
$A(\bar{D}^0\to\pi^0\pi^0)\equiv\bar{A}^{00}$ and
$A(D^-\to\pi^-\pi^0)\equiv \bar{A}^{+0}$ can be defined analogously
with $\lambda_{d,s}$ being replaced by  $\lambda^*_{d,s}$.

Note that the isospin triangle relation
\begin{equation}\label{eq:isospin_triangle}
\tfrac{1}{\sqrt{2}} A^{+-}+A^{00}=A^{+0},
\end{equation}
has been implicitly imposed by the expressions in
Eq.~\eqref{eq:isospin_rels_simp}, with a similar expression for the
conjugate modes. The amplitudes for $D\to \pi\pi$ decay modes can also
be decomposed \cite{Kamal:1987nm} in terms of the $I=0$ and $I=2$
amplitudes as follows:
\begin{align}
A^{+-}&= \sqrt{2} (A_2-A_0), \nn\\%
A^{00}&= (2 A_2+A_0), \nn\\%
A^{+0}&= 3 A_2,
\end{align}
with analogous expressions for the conjugate mode amplitudes. It may
be noted that the decay amplitudes of the conjugate modes can be
recast after rotating each of the amplitudes by the weak phase of
$\lambda_d$, which results in $A^{+0}$ and $\bar{A}^{-0}$ having the
same phase.

\begin{figure}[t]
\centering%
\includegraphics[width=\linewidth,keepaspectratio]{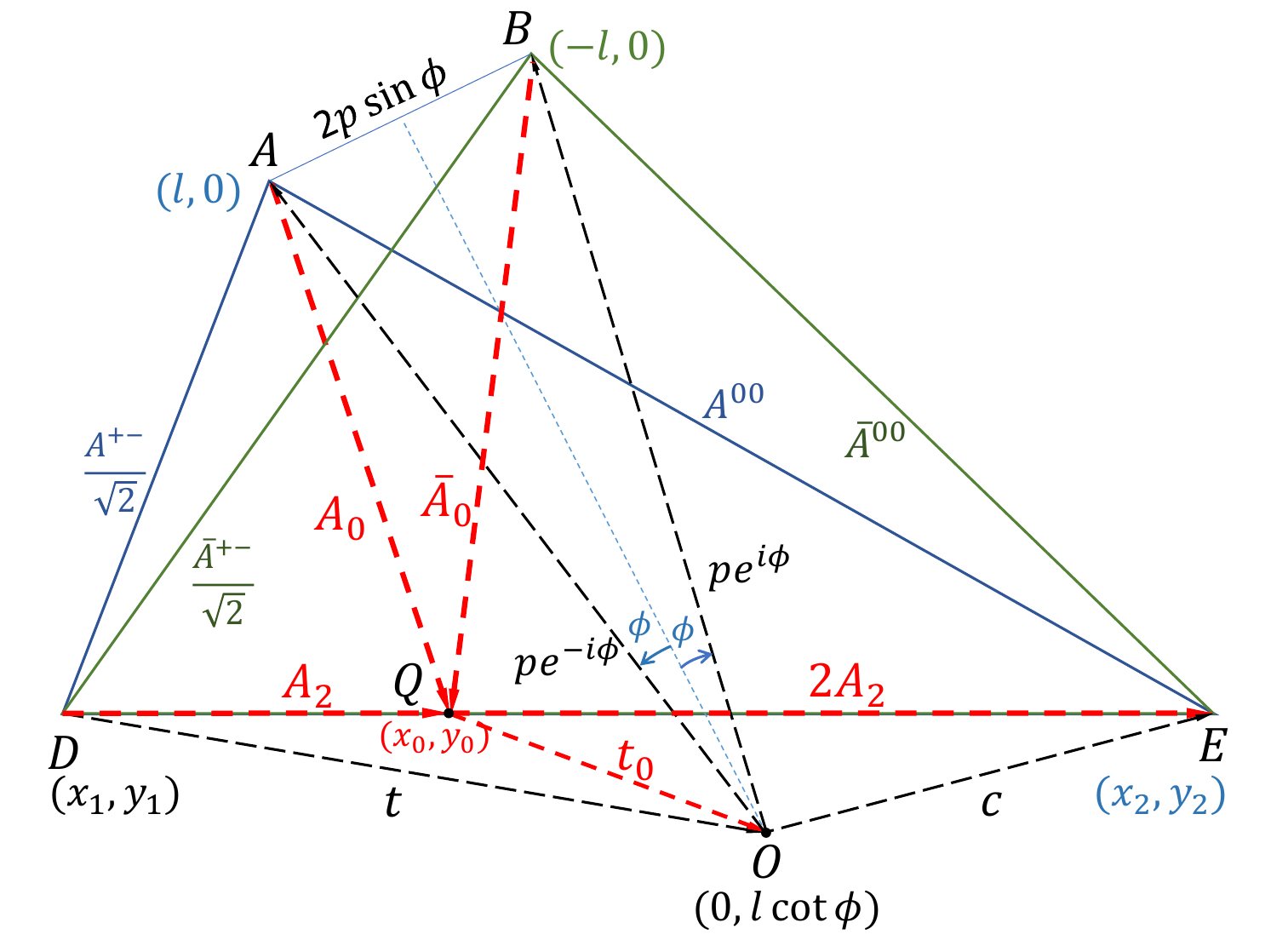}%
\caption{The isospin triangles for the $D\to \pi\pi$ and $\bar{D}\to
\pi\pi$ modes. The triangles are illustrative and do not correspond to
measured observables.}%
\label{Fig:IsospinTriangles}%
\end{figure}

In Fig.~\ref{Fig:IsospinTriangles} we show the triangles for $D\to
\pi\pi$ modes and their respective rotated conjugate modes. The
relevant information regarding the amplitudes $t$, $c$, $p$ and weak
phase $\phi$ is also encoded in the figure. Note that the triangles
are determined purely from experimental data. Given the two triangles,
the largest value of $p$ is obtained for a parametrization where
$\phi$ is the smallest. Since our aim is to determine the
size of the penguin contribution to the $D\to\pi\pi$ mode we choose
the parametrization defined in Eq.~\eqref{eq:isospin_rels_simp}. It
also turns out that the error in $\phi$ is the smallest among all
possible choices of parametrization phases that can be used to
describe our modes. This results in a more accurate measurement of
$\phi$. In Fig.~\ref{Fig:IsospinTriangles}  we also show the isospin
amplitudes $A_0$ and $A_2$ and another interesting parameter $t_0$
corresponding to the isospin zero component of $t$. The amplitude $p$
contributes only to $A_0$, whereas, $t$, $c$ contribute to both $A_2$
and $A_0$. One can express $t=t_2+t_0$ and $c=c_2+c_0$, where the
subscripts denote the isospin decomposition of $t$ and $c$. It is then
possible to relate the isospin and topological amplitudes as follows:
\begin{align}
A^{+-}&= \sqrt{2} (A_2-A_0)= \sqrt{2} (t_2+t_0+e^{i\phi}p), \nn\\%
A^{00}&= (2 A_2+A_0)=(c_2+c_0 -e^{i\phi}p), \nn\\%
A^{+0}&= 3 A_2= (t_2+c_2+t_0+c_0).
\end{align}
Using the above equations one can solve for $A_2$ and $A_0$ and easily
see that $c_0=-t_0$ and $c_2=2 t_2$ must hold so that $A_0$ and $A_2$
receives contributions only from the relevant isospin amplitudes. This
implies that $A_2 = t_2$ and $A_0=-t_0-e^{i\phi}p$, such that
\begin{align}
A^{+-}&= \sqrt{2} (A_2-A_0)= \sqrt{2} (t_2+t_0+e^{i\phi}p), \nn\\%
A^{00}&= (2 A_2+A_0)=(2 t_2-t_0 -e^{i\phi}p), \nn\\%
A^{+0}&= 3 A_2= 3t_2.
\end{align}

\begin{figure}[b]
\begin{center}
\includegraphics[width=0.8\linewidth]{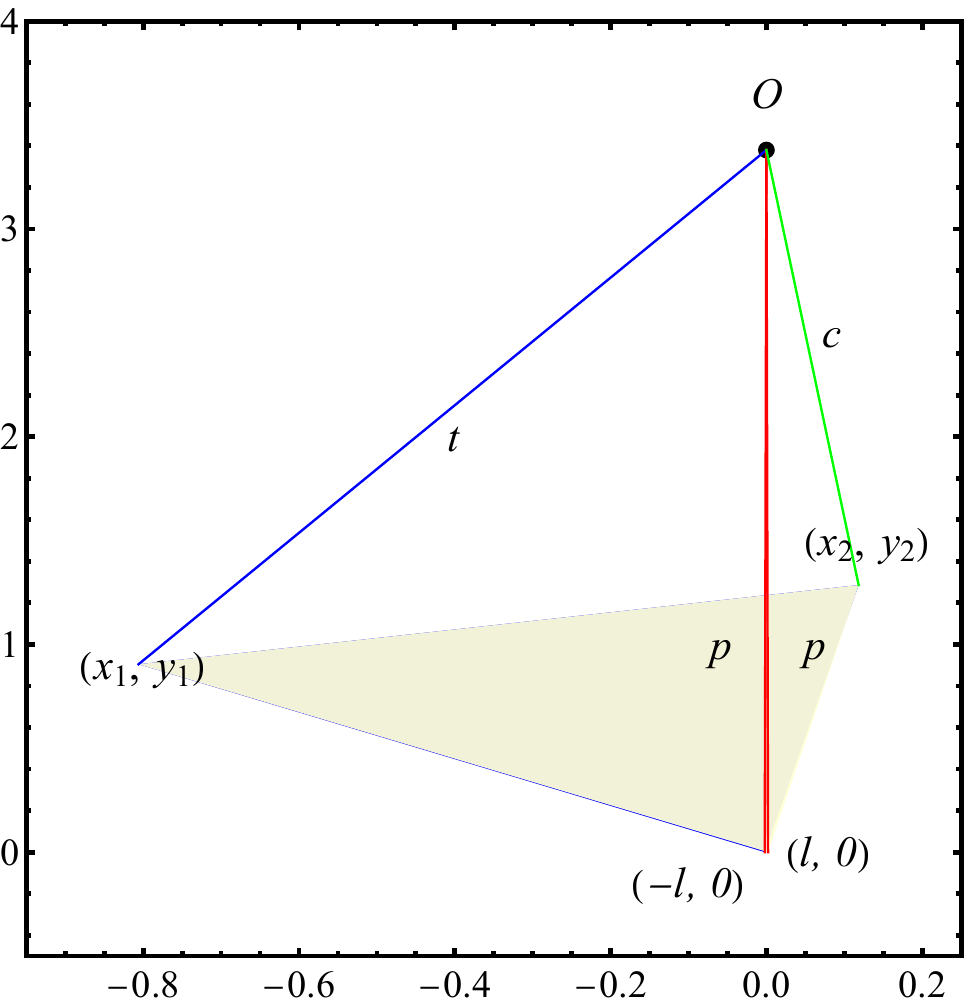}\\[5mm]%
\includegraphics[width=0.8\linewidth]{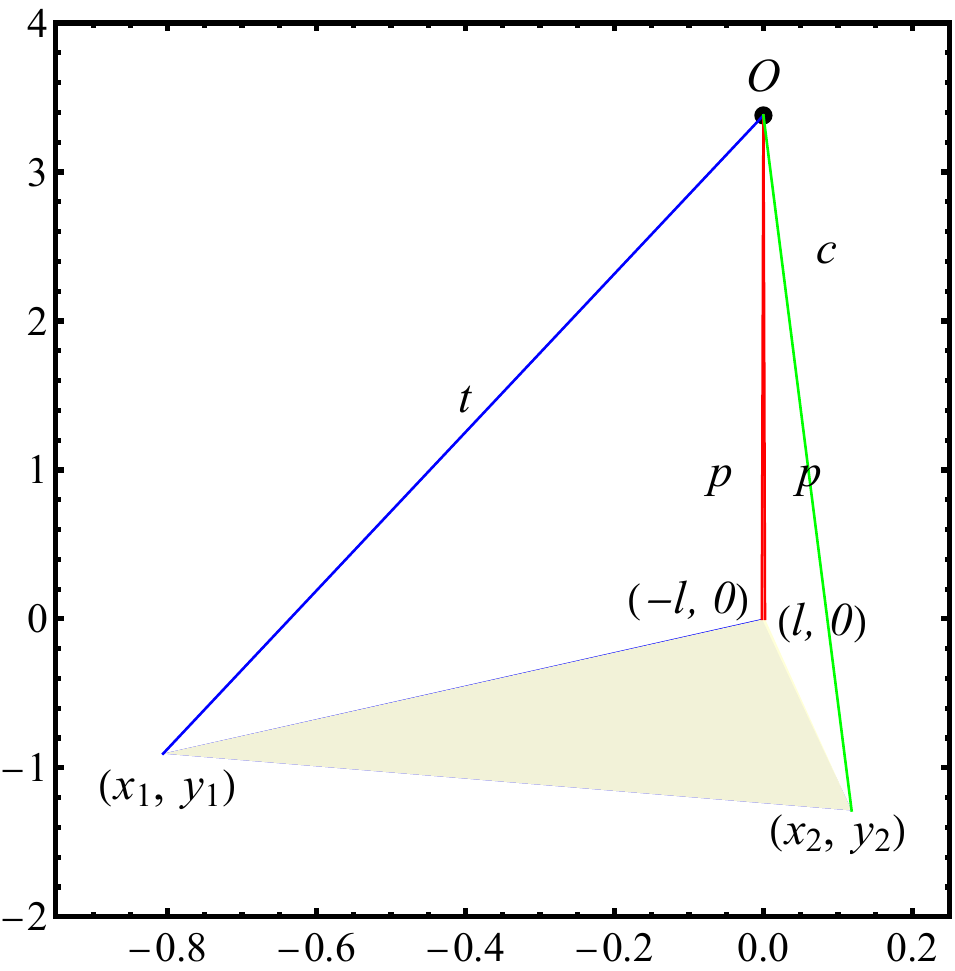}%
\end{center}
\caption{Two of the four isospin triangle orientations for the $D\to
\pi\pi$ and $\bar{D}\to \pi\pi$ modes {\em representing the central
values of the measured observables}. Note that the triangles
corresponding to the mode and conjugate mode almost overlap. The
triangle for $\bar{D}\to\pi\pi$ in dark blue is barely visible under
the translucent yellow triangle corresponding to $D\to \pi\pi$. The
triangles shown are chosen to be the ones with smaller penguin
contributions. All amplitudes are normalized to $A^{+0}$.}%
\label{Fig:Triangles}
\end{figure}

\section{Constraining the amplitudes from observables}

To determine the amplitudes in terms of the observables we assign a
coordinate system such that points $A$ and $B$ correspond to $(l,0)$
and $(-l,0)$ and the points $D$ and $E$ are $(x_1,y_1)$ and
$(x_2,y_2)$ respectively. The coordinate of $O$ is then determined to
be $(0,l\cot\phi)\equiv (0,p\cos\phi)$. The coordinate of $Q$, is
determined to be
\begin{equation}
(x_0,y_0)=\Bigl(\frac{2x_1+x_2}{3},\frac{2y_1+y_2}{3}\Bigr).
\end{equation}
The branching fractions (${\cal B}$) and direct \CP-asymmetries are
defined as,
\begin{align}
{\cal B}(D^0\to\pi^+\pi^-)&\equiv B_{\sss+-}
=\tfrac{1}{2}(\left|A^{+-}\right|^2 +\left|\bar{A}^{+-}\right|^2) ,
\nn\\%
{\cal B}(D^0\to\pi^0\pi^0)&\equiv B_{00}
=\tfrac{1}{2}(\left|A^{00}\right|^2 +\left|\bar{A}^{00}\right|^2) ,
\nn\\%
{\cal B}(D^+\to\pi^+\pi^0)&\equiv B_{+0} =\left|A^{+0}\right|^2
\frac{\tau_{D^+}}{\tau_{D^0}} ,
\end{align}
\begin{align}
a_{\sss+-}&\equiv \frac{(\left|A^{+-}\right|^2
-\left|\bar{A}^{+-}\right|^2)}{(\left|A^{+-}\right|^2
+\left|\bar{A}^{+-}\right|^2)}, \quad a_{00}\equiv
\frac{(\left|A^{00}\right|^2
-\left|\bar{A}^{00}\right|^2)}{(\left|A^{00}\right|^2
+\left|\bar{A}^{00}\right|^2)},\nn
\end{align}
where the amplitudes are defined by including the lifetime
and phase space corrections. Note that we have normalized the
amplitude $A^{+0}$ by lifetime ratios, so that all amplitudes have the
same scale. We have used PDG~\cite{ParticleDataGroup:2024cfk} and
HFLAV~\cite{HeavyFlavorAveragingGroupHFLAV:2024ctg} values
$B_{\sss+-}=(1.454\pm 0.024)\times 10^{-3}$, $B_{00}=(8.26\pm
0.25)\times 10^{-4}$ and $B_{+0}=(1.247\pm 0.033) \times 10^{-3}$. The
Belle collaboration has measured the \CP asymmetry  $a_{00}=(-3\pm
64)\times 10^{-4}$~\cite{Belle:2014evd} and recently LHCb measured the
value $a_{\sss+-}=(23.2\pm 6.1)\times 10^{-4}$\cite{LHCb:2022lry}. The
conclusions we will obtain are insensitive to the value of $a_{00}$.
We have verified that our conclusions remain unaltered even if we use
the new less restrictive value of $a_{00} = (0.30 \pm 0.72 \pm
0.20)\%$ as reported by the recent Belle~II
paper~\cite{Belle-II:2025rmf} instead. Note that both the Belle and
Belle~II measurements of $a_{00}$ are consistent with \CP symmetry in
$D^0 \to \pi^0 \pi^0$ decays.

In terms of the coordinates the observables are derived to be
\begin{align}
B_{\sss+-} = & 2 (x_1^2+y_1^2) + 2\,l^2, \qquad B_{00} = (x_2^2+y_2^2)
+ l^2, \nn\\%
B_{+0} = & (x_1^2+y_1^2) + (x_2^2+y_2^2) - 2\, x_1 x_2 - 2\,y_1 y_2,
\nn\\%
a_{\sss+-} = & -\frac{4 x_1 p\sin\phi}{B_{\sss +-}}, \qquad%
a_{00} = -\frac{2 x_2 p\sin\phi}{B_{00}}. \label{eq:obs-coord}
\end{align}
The five parameters $x_1$, $y_1$, $x_2$ $y_2$ and $l$ that define the
coordinates can be solved in terms of the five observables as follows:
\begin{align}
x_1=& -\frac{a_{\sss+-} B_{\sss+-}}{4\,l},\qquad x_2= -\frac{a_{00}
B_{00}}{2\,l},\nn\\%
y_1=& \pm\frac{\sqrt{a_{\sss +-}^2 B_{\sss +-}^2+8 B_{\sss +-}\,
l^2+16\, l^4}}{4\,l},\nn\\%
y_2=& \pm\frac{\sqrt{a_{00}^2 B_{00}^2+4 B_{00}\, l^2+ 4\,
l^4}}{2\,l},
\end{align}
and $l^2$ can now be solved upto a quadratic ambiguity using the
expression for $B_{+0}$ in Eq.~\eqref{eq:obs-coord}. The amplitudes
and strong phases of $t$, $c$, $p$, $A_2$, $A_0$, $\bar{A}_0$ and
$t_0$ can thus be determined purely in terms of experimental data.
Note that the solutions have four ambiguities that correspond to the
relative orientation of the triangles, a fact that is well known. We
show in Fig.~\ref{Fig:Triangles} the relative sizes of $t$, $c$ and
$p$ contributions (normalized to $A^{+0}$) using central values from
data. The two panels  depicted in Fig.~\ref{Fig:Triangles} correspond
to those two orientations of isospin triangles (out of four possible
orientations) that have smaller penguin contributions. The penguin
values corresponding to the the other two orientations are an order of
magnitude larger than tree contribution and are hence not considered. Note that
very large penguin contributions are indicated by data.

\begin{figure}[t]
\begin{center}
\includegraphics[width=0.9\linewidth]{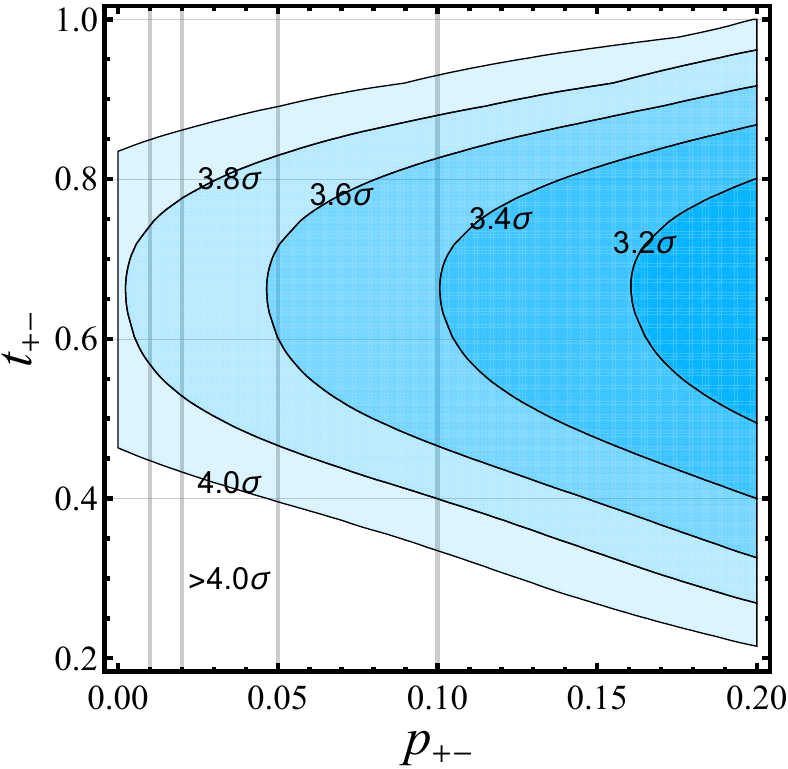}\\[5mm]%
\includegraphics[width=0.9\linewidth]{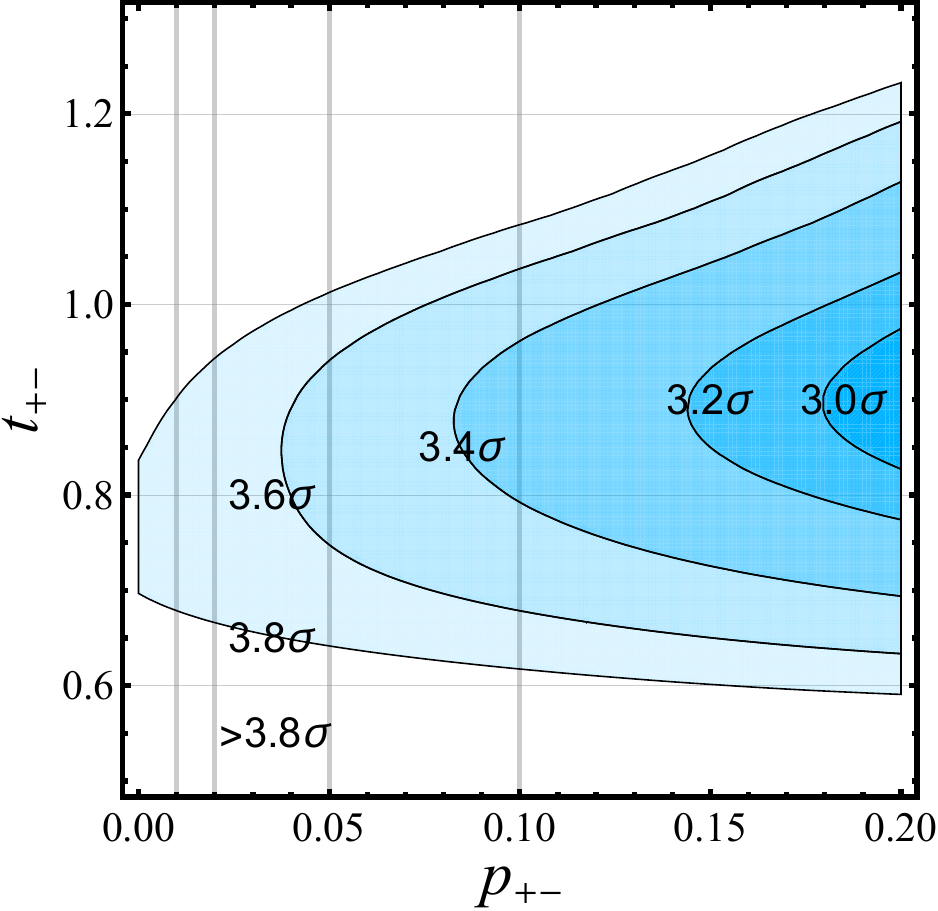}%
\end{center}
\caption{The amplitudes $p_{+-}$ versus $t_{+-}$ defined by
normalizing $p$ and $t$ by $\sqrt{B_{+-}}$, corresponding to the two
triangle orientations with smaller values of  $p$. The central values
of the fit for $(p_{+-},t_{+-})$ for the top and bottom panels  are
$(4.74,4.19)$ and $(4.80,5.43)$ respectively. It is seen that a $10\%$
value of $p_{+-}$ is at least $3\sigma$ away from the estimated value,
indicating an unacceptably large penguin contribution.}%
\label{Fig:PvsT}
\end{figure}

We write the amplitudes in terms of complex coordinate system as
follows:
\begin{align}
t= & -x_1 + i (l \cot\phi - y_1),\\%
c= & \quad x_2 + i (y_2 - l \cot\phi),\\%
p= & \qquad -i~l\csc\phi,\\%
A_0 = & (2 x_1 + x_2)/3 - l + i (2 y_1 + y_2)/3,\\%
\bar{A_0} = & (2 x_1 + x_2)/3 + l + i (2 y_1 + y_2)/3,\\%
A_2 = & (-x_1 + x_2)/3 + i  (-y_1 + y_2)/3,\\%
t_0 = & (2 x_1 + x_2)/3 + i (l \cot\phi - (2 y_1 + y_2)/3).
\end{align} 
Using complex coordinates to estimate amplitudes enables evaluation of
both magnitudes and relative phase with ease. The amplitudes of $t$,
$c$, $p$ and $t_0$ as well as their phases are evaluated using a
simulated data set of $B_{+-}$, $B_{00}$, $B_{+0}$, $a_{+-}$, $a_{00}$
and $\phi$ with a million points.

\begin{figure}[t]
\begin{center}
\includegraphics[width=0.9\linewidth]{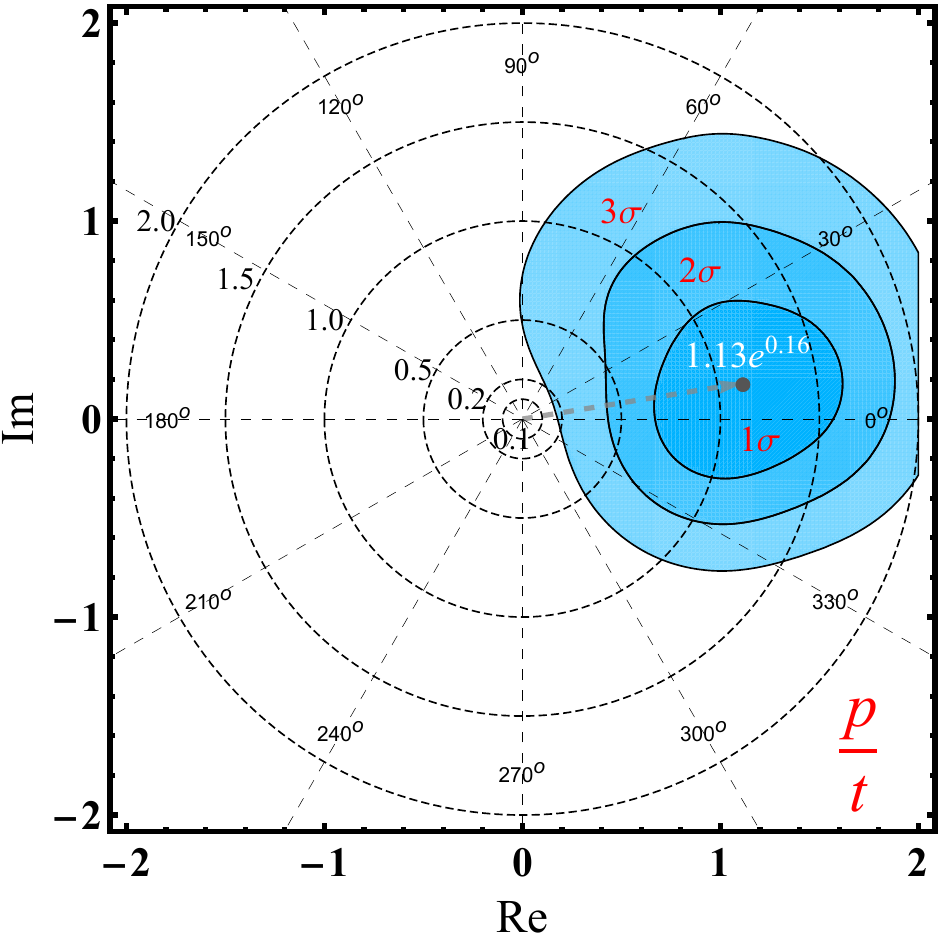}\\[5mm]%
\includegraphics[width=0.9\linewidth]{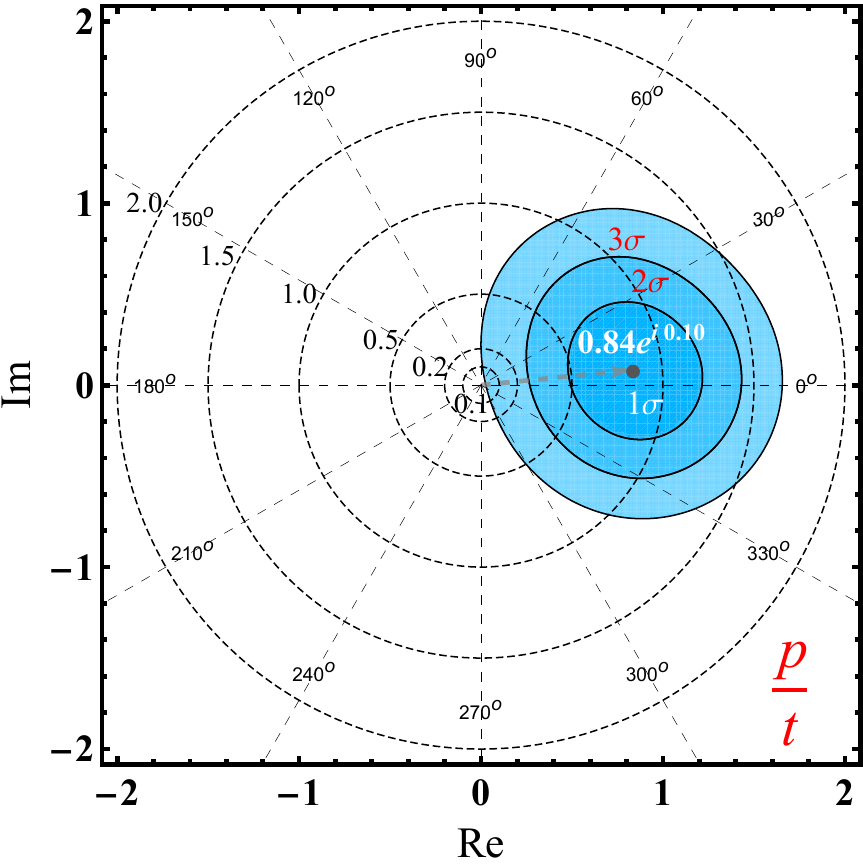}%
\end{center}
\caption{$p/t$ corresponding to the two triangle orientations smaller
$p$ are shown corresponding to the two orientations. Note that small
values of $p/t\sim 0.1$ are possible only at $\sim3\sigma$.}%
\label{Fig:PbyT}
\end{figure}

In Fig.~\ref{Fig:PvsT} the obtained normalized  amplitudes
$p_{+-}\equiv p/\sqrt{B_{+-}}$ versus $t_{+-}\equiv t/\sqrt{B_{+-}}$
are shown. The central values of the fit for $(p_{+-},t_{+-})$ are
$(4.74, 4.19)$ and $(4.80, 5.43)$, indicating that the observed
$p_{+-}$ is indeed large. If we make the choice to ignore the small
\CP violating asymmetry $a_{+-}$ relative to $A^{+-}$ and
$\bar{A}^{+-}$, we can directly compare with the estimate
$p_{\sss+-}=(0.09\pm 0.01)$ made in Ref.~\cite{Khodjamirian:2017zdu},
using light cone sum rule techniques. Similar estimates~ for
$p_{\sss+-}$ are obtained in Ref.~\cite{Lenz:2023rlq}. Since the
momentum flowing through the penguin loop are of order $1$~\gev or
larger, a perturbative calculation of this loop is not unreasonable.
Convincing arguments on the reliability of these estimates have been
made in Ref.~\cite{Nierste:2020eqb}. Note, that these are estimates of
penguin $p$ compared to $\sqrt{B^{+-}}$, and it would be unreasonable
to expect a much larger value. It is seen from Fig.~\ref{Fig:PvsT} that
a $10\%$ value of $p_{+-}$ is $\sim 3.4\sigma$ away from the estimated
value indicating an unacceptably large penguin contribution. We
note that $a_{\sss+-}$ is non-zero at $3.8\sigma$ and our results are
consistent with this measurement. Even with a simulation of a million
points no point below $p_{+-}<0.15$ was found limiting our accuracy at
vanishing $p_{+-}$ by interpolation errors. It is worth noting that
$p_{+-}\sim 1$ at $1\sigma$, $p_{+-}\sim 0.6$ at $2\sigma$,
$p_{+-}\sim 0.2$ at $3\sigma$.  Thus the data indicates a very large
penguin contribution. It may be noted that since the parametrization
of Ref.~\cite{Gavrilova:2023fzy} and our paper differ, the expressions
and estimates of the penguin also differ. However, our estimates for
$p_{+-}$ are in broad agreement with those of
Ref.~\cite{Gavrilova:2023fzy}. For completeness we have also  plotted
the $p/t$ values as a polar plot in Fig.~\ref{Fig:PbyT}. The $p/t$
values once again indicate that $p$ is indeed large and small values
of $p/t\sim 0.1$ are possible only at $\sim3\sigma$.

\begin{figure}[b]
\begin{center}
\includegraphics[width=0.9\linewidth]{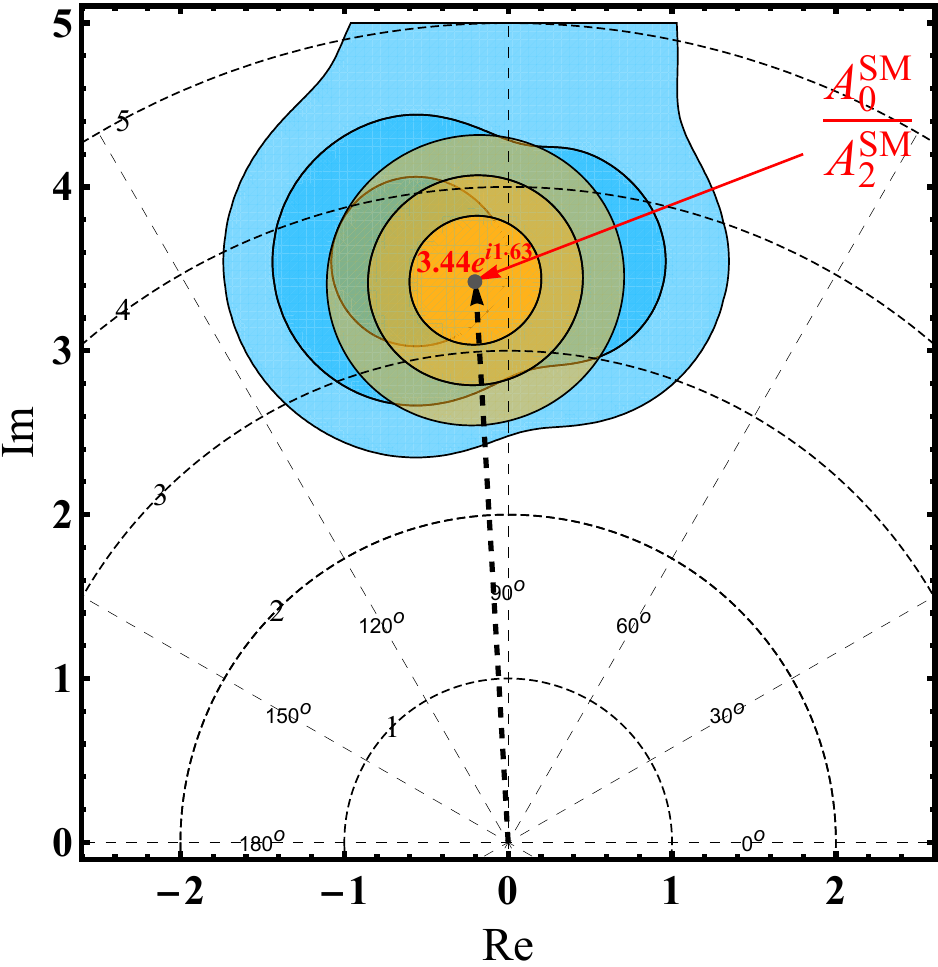}\\[5mm]%
\includegraphics[width=0.9\linewidth]{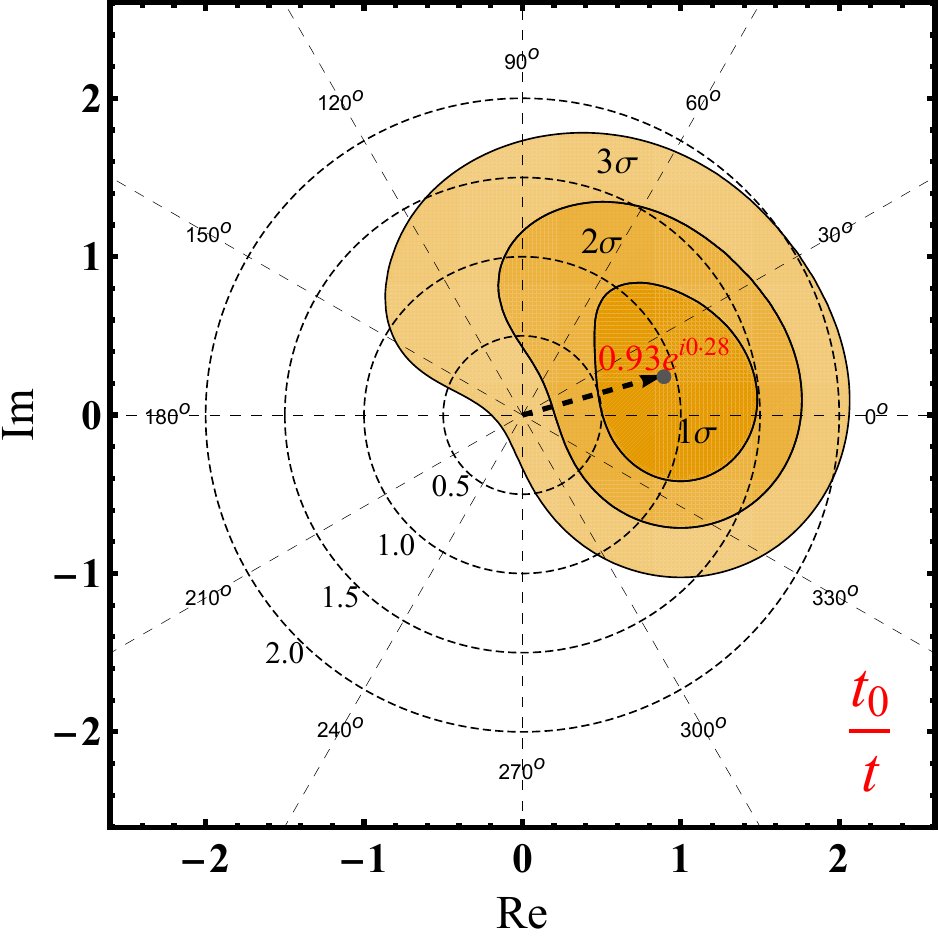}%
\end{center}
\caption{The values of $A_0/A_2$ are shown in the top panel. The
orange contours correspond to the SM and assumes $a_{+-}=0$. The blue
contours are for the observed $a_{+-}$. $|A_0/A_2|\gtrsim 2$ at
$3\sigma$ indicates  an inherent $\Delta I=1/2$ enhancement in $D\to
\pi\pi$ mode. The panel below shows contours for the ratio of $t_0/t$
indicating that $t_0$ can be almost as large as $t$.}%
\label{Fig:Isospin}
\end{figure}

In Fig.~\ref{Fig:Isospin}, the values of the ratio of isospin
amplitudes $A_0/A_2$ are shown. The orange contours referred to as SM
assumes that $a_{+-}=0$ at ${\cal O}(\lambda^4)$ as expected in the
SM. This ensures that $A_0$ and $\bar{A}_0$ amplitudes are identical. 
In the limit $a_{+-}=0$ the isospin ratio is the same for mode and
conjugate mode. The blue contours correspond to the observed $a_{+-}$.
There is another solution with the same magnitude but a phase around
$3\pi/2$, which is not shown, since it is not a true solution. The
observed $a_{+-}$ is indeed quite small, and does not result in a visible
difference in the blue contours between $D\to\pi\pi$ and $\bar{D}\to
\pi\pi$. Hence, true solutions cannot have significant deviation from
SM isospin values. The value  $|A_0/A_2|\gtrsim 2$ at $3\sigma$
irrespective of the value of the observed \CP violation, indicates an
inherent $\Delta I=1/2$ enhancement in $D\to \pi\pi$ modes in
agreement with Ref.~\cite{Franco:2012ck}. The bottom panel of
Fig.~\ref{Fig:Isospin} shows contours for the ratio of $t_0/t$, which indicates that $t_0$ can be almost as large as $t$. This indicates
that the tree topology also plays a role in  $\Delta I=1/2$
enhancement.

\section{Comments on contributions from final state interactions}

While arguments~\cite{Nierste:2020eqb} have been made on the broad
reliability of the estimates for $p_{+-}$, it may be noted that
explicit final state interactions might alter the size of the penguin
contributions. We therefore examine in a model-independent way effects
of final state interactions and draw general conclusions. In two-body
scattering of $n$-coupled channels a convenient parametrization of a
unitary S-matrix is in terms of the K-matrix,
\begin{equation}
S(s) = (1 - i K(s))^{-1} (1 + i K(s)),
\end{equation}
where $K(s)$ is a $n\times n$ hermitian matrix and $s$ is set equal
to $m_D^2$ for $D$-decays. A set of $n$ un-unitarized weak decay
amplitudes $A^{\sss O}(s)$ are unitarized into a set $A^{\sss U}(s)$
through multichannel final state interactions via a matrix equation,
\begin{equation}
A^{\sss U}(s)=(1 - i K(s))^{-1} A^{\sss O}(s).\label{eq:Unitarity-1}
\end{equation}
Since strong interactions conserve isospin and $G$-parity,
Eq.~\eqref{eq:Unitarity-1}  has to be written for each isospin and
G-parity (where it can be defined). Let $U$ be a unitary matrix that
diagonalizes the Hermitian matrix $K$ with real eigenvalues
$\lambda_i~(i=1, \cdots,n)$. Then,
\begin{equation}
\sum_{i}^{n} A_i^{{\sss U}\dagger} A^{\sss U}_i=\sum_{i}^{n} 
(U A^{{\sss O}})^\dagger_i (1+\lambda_i^2)^{-1} (U A^{\sss O})_i.
\end{equation}
Since $\lambda_i$ are real this results in the inequality
\begin{equation}
\sum_{i}^{n} \left|A^{\sss U}_i\right|^2 \le \sum_{i}^{n}
\left| A^{\sss O}_i \right|^2 . \label{eq:Unitarity_const}
\end{equation}
This inequality implies that the sum of transition probabilities after
inter-channel mixing cannot exceed the sum before mixing. Clearly all
channel contributing to any specific isospin channel can re-scatter
among each other. The channels $D^0\to\pi \pi$, $D^0\to K K$ as well
as other channels can re-scatter in the isospin zero channel resulting
in the huge enhancement in the $I=0$, $D\to \pi\pi$ amplitudes. It has
to be kept in mind however, that since the penguin amplitudes $p$ has
a weak phase it cannot receive contributions from any other channel
that does not have the same weak phase. {\em Eq.\eqref{eq:Unitarity-1}
must therefore apply separately to the $p$ amplitudes}, implying the
constraint:
\begin{equation}
\sum_{i=\pi\pi, KK,\cdots}^{}\left|p^{\sss U}_i\right|^2 \le
\sum_{i=\pi\pi, KK,\cdots}^{}\left|p^{\sss O}_i\right|^2,
\end{equation}
where, $\cdots$ represent any other channel contributing to $I=0$. If
the anomalously large penguin contributions arises from within the SM,
one must generate large penguin contributions in at least one of the
channels that re-scatters to produce the observed asymmetries. This is
difficult to accommodate within the SM. If $D\to KK$ penguin amplitude
re-scatters to $D\to \pi\pi$ such that $D\to \pi\pi$ penguin
contributions become very large it would result in $D\to KK$ penguin
amplitude becomes very small and consequently a vanishing \CP
asymmetry. For any other channel~\cite{Franco:2012ck} contributing to
$I=0$, the amplitude of the penguin must be established to be large
theoretically. We must therefore conclude that attempts
\cite{Bediaga:2022sxw} to explain large penguins contributions and
large \CP asymmetry, may inherently violate unitarity in re-scattering
and should be viewed with skepticism.

In our analysis we have ignored electroweak penguins ($p_{\sss EW}$)
which is justifiable within SM, but they can be sizable in the
presence of NP. In the absence of $p_{EW}$ the measurements of
branching ratios and the two \CP asymmetries provide a complete
determination of all parameters that describe the decay. However, this
is not true in the presence of $p_{EW}$ where one parameter remains
unsolved~\cite{Nayak:2020oba} without an additional input. The direct
\CP asymmetry measured in the $D^+\to \pi^+\pi^0$ mode is
$a_{+0}=(4.0\pm 8.0)\times 10^{-3}$. Hence, the two triangles will not
in principle share a common base, since $A^{+0}$ and $\bar{A}^{+0}$
can differ by as much as $0.012$ at $1\sigma$ confidence\footnote{This
is the precision with which the presence of NP should explain the
observation.}. However, this difference is small given the estimate of
penguin contributions shown in Fig~\ref{Fig:PvsT}, hence to a very
good approximation we can assume that $p$ includes $p_{EW}$
contributions in all our estimations. We note, that even a small
$p_{EW}=0.005$ and new physics weak phase of $\pi/6$ can produce an
angle between $A^{+0}$ and $\bar{A}^{+0}$ of $8.2\deg$. However, this
would not effect our analysis.

\section{Probing the size of generic NP contribution}

We next examine if NP can resolve the puzzle of large penguin and the
observed direct \CP asymmetry. This has been a subject of intense
study and it is generally believed that a variety
\cite{Grossman:2006jg, Dery:2019ysp, Chala:2019fdb, Hiller:2012wf,
Calibbi:2013mka, Schacht:2022kuj} of NP scenarios can contribute to
the observed anomaly. Let us assume that NP also contributes to the
decay mode via the $p+p_{EW}$ topologies. We parameterize such NP by
$N e^{i\phi_{\sss N\!P}}$, where $\phi_{\sss N\!P}$ is the weak phase
of NP and $N$ includes the magnitude and any relevant strong phase.
This NP amplitude can easily be rewritten such that $N e^{i\phi_{\sss
N\!P}}\equiv N_1+N_2  e^{i\phi}$.  NP effects the amplitudes such that
\begin{align}
\label{eq:rels_simp2} A^{+-}&\to \sqrt{2} ( t + e^{i\phi} p+N
e^{i\phi_{\sss N\!P}})\nn\\%
&= \sqrt{2} \Big( (t+N_1)+(p+N_2)e^{i\phi} \Big), \nn\\%
A^{00}&\to  c- (e^{i\phi} p+N e^{i\phi_{\sss N\!P}})\nn\\%
&= (c-N_1)-(p+N_2)e^{i\phi}, \nn\\%
A^{+0}&= t+c.
\end{align}
Note that the amplitudes with NP contributions still obey the isospin
triangle relation of Eq.~\eqref{eq:isospin_triangle}.

\begin{figure}[!h]
\centering\includegraphics[width=0.8\linewidth]{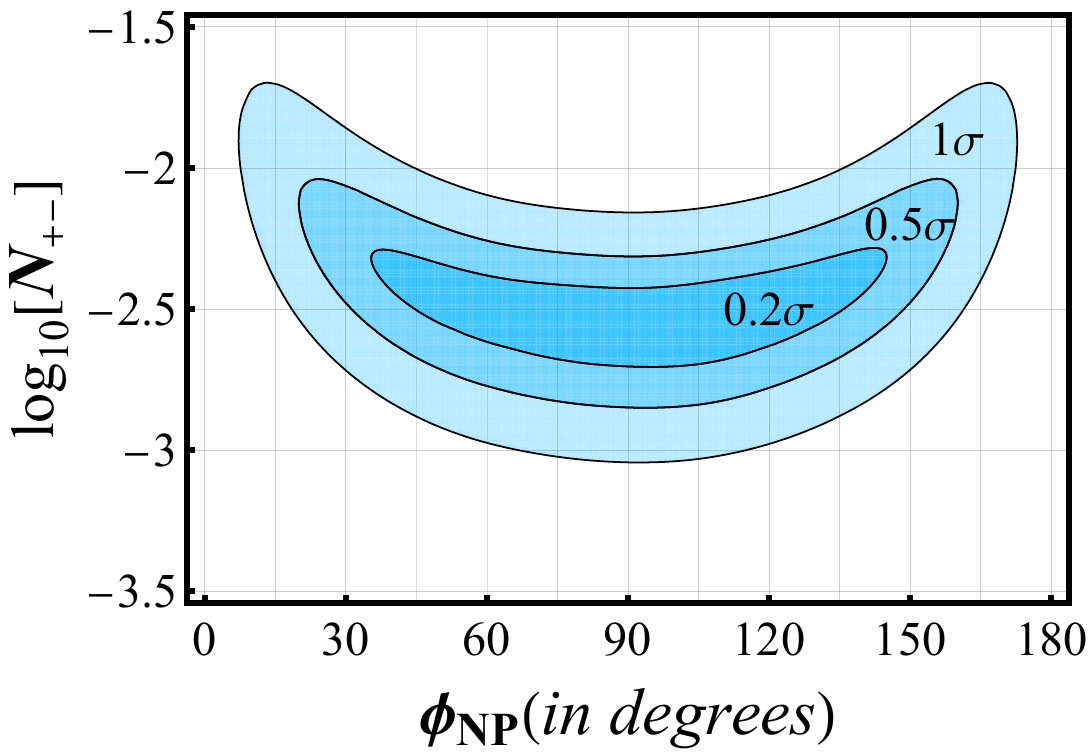}%
\caption{Contours for $\log_{10}[N_{+-}]$ versus $\phi_{\sss N\!P}$
(in degrees). It is easy to see that reasonable value of $N_{+-}$
smaller than $p^{\rm\sss SM}_{+-}$ can result in large observed
penguin and \CP violation.}%
\label{Fig:NPvalues}
\end{figure}

It is easy to see that $N_1$ and $N_2$ can be expressed in terms of
$N$, $\phi_{\sss N\!P}$ and $\phi$ to be
\begin{align}
N_1= & N \cos\phi_{\sss N\!P} - N \cot\phi \sin\phi_{\sss N\!P},\nn\\%
N_2= & N \csc\phi \sin\phi_{\sss N\!P}.\label{eq:NP}
\end{align}
Clearly taking into account NP results in $t$ being modified by $N_1$
and $p$ by $N_2$, so that the weak phases are aligned. This implies
that both $t$ and $p$ are altered by NP. The observed value of the
penguin  would include contributions from both SM penguin  and NP.
Hence,
\begin{equation}\label{eq:pbyt_obs}
p^{\rm\sss obs}_{+-}= (p^{\rm\sss SM}_{+-} + N_{+-} e^{i\delta_{NP}}
\csc\phi \sin\phi_{\sss N\!P}) e^{i\phi},
\end{equation}
where all amplitudes are normalized by $\modulus{A_{+-}}$; $p^{\rm\sss
obs}_{+-}\equiv p^{\rm\sss obs}/\modulus{A_{+-}}$, $p^{\rm\sss
SM}_{+-}\equiv p^{\rm\sss SM}/\modulus{A_{+-}}$,
$N_{+-}=N/\modulus{A_{+-}}$ and $\delta_{NP}$ is the difference
between the strong phase of $N$ and $p$. Without loss of generality,
the domain of $\delta_{NP}$ is chosen such that the amplitudes
$p^{\rm\sss SM}_{+-}$ and $N_{+-}$ are always positive. We hence have
\begin{align}\label{eq:NP-sol} 
(p^{\rm\sss obs}_{+-})^2= &(p^{\rm\sss SM}_{+-})^2+2 N_{+-} p^{\rm\sss
SM}_{+-} \cos\delta_{NP}\csc\phi \sin\phi_{\sss N\!P} \nn\\%
&~~~\qquad + N_{+-}^2 \csc^2\!\phi \sin^2\!\phi_{\sss N\!P}.
\end{align}
Note that $p^{\rm\sss obs}_{+-}$ has already been obtained  from data
and $p^{\rm\sss SM}_{+-}$ has been estimated in
Ref.~\cite{Khodjamirian:2017zdu}. We can use Eq.~\ref{eq:NP-sol} to
solve for $N_{+-}$ in terms of $\phi_{\sss N\!P}$ and $\delta_{NP}$,
since $\phi$ is known with errors. The simulated data set used to
obtain $p^{\rm\sss obs}_{+-}$ is extended to include randomly
generated Gaussian values of $p^{\rm\sss SM}_{+-}$, whereas
$\delta_{NP}$ and $\phi_{\sss N\!P}$ are randomly varied in the range
$(0,\pi)$. The range for $N_{+-}$ as function of $\phi_{\sss N\!P}$ is
shown in Fig.~\ref{Fig:NPvalues}. We find that large \CP violation
observed and resulting very large penguin can arise due to a $N_{+-}$
with a larger weak phase, even with $N_{+-}$  smaller than $p^{\rm\sss
SM}_{+-}$, indicating that NP could result in the larger than expected
\CP asymmetry. The reader may wonder how a small NP amplitude can
generate a large \CP asymmetry. We emphasize that even a small
$\phi_{\sss N\!P}=\pi/6$ results in an enhancement of $\sim 800$ in
$a_{\sss +-}$ compared to that from SM phase $\phi$, resulting in a
much larger asymmetry even with a smaller $N_{+-}$ compared to SM
penguin $p_{+-}$.

While the case for NP looks robust, it may be noted that it depends
largely on the measured value of $a_{\sss+-}^{KK}=(7.7\pm 5.7)\times
10^{-4}$~\cite{LHCb:2022lry}.  Let us consider a scenario, in future
measurements, where $a_{\sss+-}^{\sss KK}=(-7.6\pm  2.3)\times
10^{-4}$ indicating a shift by $2.5\sigma$. The robust measurement of
the difference between direct \CP asymmetries , $\Delta
a_{CP}=a_{\sss+-}^{KK}-a_{\sss+-}^{\pi\pi}=(-15.4\pm 2.9) \times
10^{-4}$~\cite{LHCb:2019hro}, would then imply
$a_{\sss+-}^{\pi\pi}\equiv a_{\sss+-}=(7.8\pm 2.6)\times 10^{-4}$, a
downwards shift of $2.5\sigma$. The extracted value of $p_{+-}$ then
reduces to within $\sim1\sigma$ of the estimate
$p_{+-}=0.09$~\cite{Khodjamirian:2017zdu}, indicating consistency with
SM and no evidence for enhanced penguins. Interestingly, in such a
case one would also have an almost exact $U$-spin symmetry.

\section{Conclusion}

The evidence of larger than expected \CP violation in singly Cabibbo
suppressed D meson decay $D\to \pi^+\pi^-$ has been keenly debated in
literature with the view of ascertaining if the observed \CP violating
asymmetry is indeed a signal of NP. We avoid the difficulties in
arriving at a convincing conclusion by providing an estimate of the
size of the penguin contributions directly from experimental data.
Secondly, we  examine final state interactions in a model-independent
way using only unitarity, in contrast to previous approaches. The
model-independent constraints derived indicate that attempts
\cite{Bediaga:2022sxw} to explain large penguin contributions and
large \CP asymmetry, may inherently violate unitarity in re-scattering
and should be viewed with skepticism. We conclude that the measurement of
\CP violation in $D\to \pi^+\pi^-$, $a_{\sss+-}=(23.2\pm 6.1)\times
10^{-4}$, provides evidence of NP at about $3\sigma$
significance. In a model-independent approach we show how a very small
contribution from NP with a large weak phase alleviates the problem.
If this evidence is indeed a signal of NP at the level observed in
$D\to \pi^+\pi^-$, large CP violation must show up in other modes and
experiments. Searches for other signals of \CP violation in $D$ decays
are needed to substantiate the evidence of NP.

\acknowledgments 
T.E.B. thanks the DOE Office of High Energy Physics for support
through DOE grant DE-SC0010504.

\end{document}